\documentclass[aps,prd,nofootinbib,showpacs,twocolumn,floatfix,superscriptaddress]{revtex4}
\usepackage{amssymb,amsfonts}
\usepackage{bm} % bold math
\usepackage{graphicx}
\newcommand{\w}[1]{\bm{#1}}
\newcommand{\be}{\begin{equation}}
\newcommand{\ee}{\end{equation}}
\newcommand{\bea}{\begin{eqnarray}}
\newcommand{\eea}{\end{eqnarray}}

\newcommand{\el}{\w{\ell}}

\newcommand{\tD}{\tilde D}
\newcommand{\tDS}{{}^{2}\!\tilde D}
\begin{document}

\title{Isolated Horizon structures in quasiequilibrium black hole initial data}

\newcommand*{\IAA}{Instituto de Astrof\'{\i}sica de Andaluc\'{\i}a, CSIC, Apartado Postal 3004, Granada 
        18080, Spain} 
\newcommand*{\MEU}{Laboratoire Univers et Th\'eories (LUTH), Observatoire de Paris, CNRS, 
Universit\'e Paris Diderot ; 5 place Jules Janssen, 92190 Meudon, France}

\author{Jos\'e Luis Jaramillo}
\email{jarama@iaa.es} 
\affiliation{\IAA}
\affiliation{\MEU}

\date{24 April 2009}

\begin{abstract}
Isolated horizon conditions 
enforce the time invariance of both the intrinsic and the extrinsic geometry of
a (quasilocal) black hole horizon.
Nonexpanding horizons, only requiring the invariance of the intrinsic 
geometry, 
have been successfully employed in the (excision) initial data
of black holes in instantaneous equilibrium. 
Here we propose the use of the full isolated horizon structure 
when solving the elliptic system resulting from the 
complete set of conformal 3+1 Einstein equations 
under a quasiequilibrium ansatz  prescription. 
We argue that a set of geometric inner boundary conditions for this 
extended elliptic system then follows,
determining the shape of the excision surface.

\end{abstract}


\maketitle

%%%%%%%%%%%%%%%%%%%%%%%%%%%%%%%%%%%%%%%%%%%%%%%%%%%%%%%%%%%%%%%%%%%%%%%%%%%%%%%
\section{General problem and specific goal}
\label{s:Gen}

Here we discuss the use of the whole structure of an isolated horizon 
in order to set inner boundary conditions for constructing 
quasiequilibrium (binary) black hole initial data,
when solving the full set of 
Einstein equations under a quasiequilibrium approximation.
This extends existing proposals in the literature that exploit
partially the isolated horizon framework, specifically the so-called
{\em nonexpanding horizon} (NEH) structure.  
We focus on approaches to initial data based on a 
conformal decomposition of the data 
on a spacelike slice $\Sigma$.
Part of the data  $(\gamma_{ab},K^{ab})$,
respectively the 3-metric and the extrinsic curvature,
must be chosen {\em a priori} to set the constraints
as a well-posed elliptic system. The physical content encoded in that choice 
is difficult to assess, and this has led to
schemes where the whole set of Einstein 
equations is solved under a certain quasi-equilibrium approximation 
(e.g. \cite{ShiUryFri04,BonGouGra04,CooBau08}). 
We consider black hole data implementing an excision technique.
Inner boundary conditions must then be discussed to complete the elliptic 
system. 
In the setting of the (extended) conformal thin sandwich (XCTS) 
equations \cite{CTS}, 
inner boundary conditions that characterize
the excised sphere as 
an apparent horizon 
in instantaneous equilibrium have been presented in \cite{Cook_etal,IH_Meudon}.
These boundary conditions exploit 
the nonexpanding horizon notion, 
the first level in the hierarchy considered in the isolated horizon 
framework \cite{AshKri04,GouJar06}. 
The specific purpose of this paper is to argue
that,  going beyond nonexpanding horizons and using 
the whole (strongly) isolated horizon structure,
inner boundary conditions for the (conformal) metric
can be derived. These conditions can then be implemented in 
the resolution of the
full set of Einstein equations under a quasiequilibrium 
ansatz \cite{ShiUryFri04,BonGouGra04,CooBau08}, where they 
determine the extrinsic curvature
(shape) of the excision surface. 

As a further physical motivation for the full isolated horizon
structure, we comment on a one-parameter family of horizon slicings with potential
interest in the discussion of quasilocal black hole linear momentum.

\label{s:notation}
\subsection{3+1 decompositions.} 
We consider 3+1 foliations $\{\Sigma_t\}$ of
spacetimes  $({\cal M}, g_{ab})$, with Levi-Civita connection
$\nabla_a$. We denote by $n^a$
the timelike unit normal to the spacelike $\Sigma_t$.
The  evolution vector is 
$t^a=N n^a+\beta^a$, with $N$ the lapse function and
$\beta^a$ the shift vector.
We denote by  $\gamma_{ab}$ the induced metric on $\Sigma_t$,
i.e. $\gamma_{ab}=g_{ab}+n_a n_b$, and  by $D_a$ its
associated Levi-Civita connection.
The sign convention for the extrinsic curvature of $(\Sigma_t, \gamma_{ab})$ 
inside $({\cal M}, g_{ab})$
is $K_{ab}:=-\frac{1}{2}{\cal L}_n \gamma_{ab}=
-{\gamma_a}^c \nabla_{c}n_b$.

\subsection{Closed 2-surfaces.}
Given a 2-surface ${\cal S}_t \subset \Sigma_t$, $s^a$ is the spacelike unit 
normal vector  
pointing {\em outwards} (towards infinity in the asymptotically flat case) and tangent to $\Sigma_t$.
The normal plane $T_p^\perp {\cal S}_t$ at 
$p\in {\cal S}_t$ is spanned by 
$n^a$ and $s^a$.
Alternatively, one can span  $T_p^\perp {\cal S}_t$ 
in terms of the {\em outgoing} and  {\em ingoing} null normals, respectively
denoted as $\ell^a$ and $k^a$, satisfying $k^c \ell_c = -1$.
Directions defined by $\ell^a$ and $k^a$  are uniquely determined,
but a boost-normalization freedom remains. We can then write
$\ell^a = f\cdot(n^a + s^a)$ and 
$k^a = \frac{1}{2f}(n^a - s^a)$, up to a factor $f$.
The induced metric on ${\cal S}_t$ is given by: $q_{ab}=g_{ab}+k_a \ell_b
+ \ell_a k_b = g_{ab} +n_a n_b - s_a s_b = \gamma_{ab} - s_a s_b$,
The Levi-Civita connection associated with $q_{ab}$ is
${}^2\!D_a$.
We define the {\em second fundamental tensor} of $({\cal S}, q_{ab})$ in 
$({\cal M}, g_{ab})$
as ${\cal K}^c_{ab}:= {q^d}_a {q^e}_b 
\nabla_d {q^c}_e $, so that 
\bea
\label{e:2nd_fund_form}
 {\cal K}^c_{ab}= n^c \Theta^{(n)}_{ab} + 
s^c \Theta^{(s)}_{ab} =  k^c \Theta^{(\ell)}_{ab}+
 \ell^c \Theta^{(k)}_{ab} \ ,
\eea
where the {\em deformation tensor}
$\Theta^{(v)}_{ab}$ associated with $v^a$ normal to ${\cal S}$ is 
defined as $\Theta^{(v)}_{ab}:=  {q^c}_a {q^e}_b \nabla_c v_d$.
The extrinsic curvature $H_{ab}$ of $({\cal S}_t, q_{ab})$ 
in $(\Sigma_t, \gamma_{ab})$
corresponds to $\Theta^{(s)}_{ab}$: 
$ H_{ab}\equiv {q^c}_a {q^d}_b D_c s_d =  \Theta^{(s)}_{ab}$.
Extrinsic curvature information of $({\cal S}, q_{ab})$ 
in $({\cal M}, g_{ab})$ is completed by the {\em normal fundamental forms}
associated with normal vectors $v^a$. 
We shall employ the 1-form $\Omega^{(\ell)}_a := k^c {q^d}_a \nabla_d \ell_c$.

\medskip 

\section{Isolated Horizons}
\label{s:IH}

Isolated horizons \cite{AshKri04} provide a 
setting for black hole horizons
in equilibrium inside a dynamical spacetime. 
The minimal notion of equilibrium is given by a 
NEH. A NEH is a $S^2 \times \mathbb{R}$ null
hypersurface $\mathcal{H}$, 
on which the Einstein equation holds under a certain energy condition, and that
is sliced by marginally (outer) trapped surfaces. That is, the expansion associated 
with $\ell^a$ vanishes on $\mathcal{H}$:
$\theta^{(\ell)}=q^{ab}\Theta^{(\ell)}_{ab} = 0$.
The geometry of a NEH is characterized by the pair $(q_{ab}, \hat{\nabla}_c)$, where
$q_{ab}$ is the induced null metric on ${\cal H}$ and $\hat{\nabla}_a$ is the
unique connection (not a Levi-Civita one) induced from the ambient spacetime
connection. 
The connection  $\hat{\nabla}_a$ characterizes the {\em extrinsic geometry} of the NEH
tube. 
Expression $\hat{\nabla}_a \ell^b = \omega^{(\ell)}_a \ell^b$ defines a
1-form $\omega^{(\ell)}_a$ intrinsic to  ${\cal H}$, for each $\ell^a$.
A {\em surface gravity} $\kappa_{(\ell)}$ is then 
defined from the acceleration of $\ell^a$, i.e. 
$\ell^c\hat{\nabla}_c \ell^a = \kappa_{(\ell)}\ell^a$.
In terms of the normal fundamental form introduced above, we can write
$\omega^{(\ell)}_a = \Omega^{(\ell)}_a - \kappa_{(\ell)} k_a $ 
(cf. Eq. (5.35) in \cite{GouJar06}). 

A hierarchy on $\cal{H}$ results from the progressive demand
of geometry invariance under
the $\ell^a$ (evolution) flow:

\noindent (i) A NEH is characterized by the {\em time invariance} of the
intrinsic geometry $q_{ab}$: ${\cal L}_\ell q_{ab} = 2 \; \Theta^{(\ell)}_{ab} = 0$.

\noindent (ii) A weakly isolated horizon (WIH) is a NEH, together with 
an equivalence class of null normals $[\ell^a]$, for which the 1-form $\omega^{(\ell)}_a$ 
is time-invariant: ${\cal L}_\ell \omega^{(\ell)}_a = 0$. This is equivalent
to the 
surface gravity constancy:
$\hat{\nabla}_a \kappa_{(\ell)} = 0$.

\noindent (iii) An isolated horizon (IH) is a WIH on which the whole extrinsic
geometry is time-invariant: $[{\cal L}^{\cal H}_\ell, \hat{\nabla}_a]=0$.

NEH and IH equilibrium levels represent genuine restrictions
on the geometry of ${\cal H}$ as a hypersurface in $({\cal M}, g_{ab})$
[this is not the case for a
WIH, that can always be constructed on a NEH by an appropriate choice 
of $\ell^a$]. 
Writing $\Theta^{(\ell)}_{ab} = \sigma^{(\ell)}_{ab} + \frac{1}{2} \theta^{(\ell)} q_{ab}$,
NEH conditions translate into the vanishing of the shear $\sigma^{(\ell)}_{ab}$
and expansion $\theta^{(\ell)}$:
\bea
\label{e:NEH_BC}
\Theta^{(\ell)}_{ab} = 0 \ \ \ \Longleftrightarrow \ \ \ \theta^{(\ell)} = 0 \ \ , 
\ \ \sigma^{(\ell)}_{ab}=0 \ .
\eea
These NEH conditions fix part of the extrinsic curvature of ${\cal S}_t$ 
[cf. Equation (\ref{e:2nd_fund_form})] 
and have been considered in \cite{Cook_etal,IH_Meudon}.
In the next equilibrium level,  (strongly) IH conditions 
can be expressed as (cf. Eq. (5.2) in \cite{AshBeeLew02} or  Eq. (9.4) in \cite{GouJar06})
\bea
\label{e:IHconstraint}
\kappa_{(\ell)} \Theta^{(k)}_{ab}&=&
   \frac{1}{2} \left( {}^2\!D_a \Omega^{(\ell)}_b + {}^2\!D_b \Omega^{(\ell)}_a \right) 
+ \Omega^{(\ell)}_a \Omega^{(\ell)}_b 
\nonumber \\
  && - \frac{1}{2}  {}^2\!R_{ab} 
   + 4\pi \left({q^c}_a {q^d}_b T_{cd} - \frac{T}{2} q_{ab}\right) \ ,
\eea  
where ${}^2\!R_{ab}$ is the $q_{ab}$-Ricci tensor, $T_{ab}$ is the stress-energy 
tensor,
and $T=g^{cd}T_{cd}$.
Condition (\ref{e:IHconstraint}) represents a geometric
constraint on the IH data (cf. discussion in \cite{AshBeeLew02}). 
From Eq. (\ref{e:2nd_fund_form}), 
NEH conditions (\ref{e:NEH_BC}) together 
with the IH constraint (\ref{e:IHconstraint}) fix completely the second fundamental form 
${\cal K}^c_{ab}$ of ${\cal S}_t$.
Surface gravity $\kappa_{(\ell)}$ in (\ref{e:IHconstraint})
must be set to a constant value, $\kappa_{(\ell)}=\kappa_o$. 
This entails no loss of generality, due to the gauge freedom in the 
WIH structure \cite{AshKri04,GouJar06}.
Inserting in (\ref{e:IHconstraint}) the 3+1 expressions of $\Omega^{(\ell)}_a$ 
and $\Theta^{(k)}_{ab}$
\bea
\label{e:OmegakappaThetak}
\Omega^{(\ell)}_a &=& - {q^c}_a s^d K_{cd} + {}^2D_a \mathrm{ln}f    \nonumber \\
 \Theta^{(k)}_{ab} &=& - \frac{1}{2f}\left(H_{ab} + {q^c}_a {q^d}_b K_{cd}  \right) \ \ ,
\eea
a constraint on the 3+1 fields evaluated on ${\cal S}_t$ follows. 
From the trace of (\ref{e:IHconstraint}), expansion
$\theta^{(k)} = \Theta^{(k)}_{cd}q^{cd}$ satisfies
\bea
\label{e:thetak}
\theta^{(k)} = \frac{1}{\kappa_o}\left(  {}^2\!D^c \Omega^{(\ell)}_c  +   
\Omega^{(\ell)}_c \Omega_{(\ell)}^c -\frac{1}{2} {}^2\!R
+    4\pi \left(q^{cd} T_{cd} - T \right) \right) .
\eea
Adapting $\ell^a=f\cdot(n^a + s^a)$ to the 3+1 slicing, 
i.e. $f=N$, Eq. (\ref{e:thetak})
becomes the condition (11.24) in \cite{GouJar06} 
for the lapse $N$. 
The traceless part of (\ref{e:IHconstraint}) completely fixes
the ingoing shear $\sigma^{(k)}_{ab}$.  From the equality between 
the two independent
expressions from (\ref{e:IHconstraint}) and (\ref{e:OmegakappaThetak}) for
\bea
\label{e:sigmak} 
\sigma^{(k)}_{ab} &=&  \Theta^{(k)}_{ab} - \frac{1}{2} \theta^{(k)} q_{ab} \ \ ,
\eea
it follows a geometric condition on 2 degrees of freedom associated with the
traceless part of the extrinsic curvature of the excised surface ${\cal S}$ 
as embedded in $\Sigma$. 

\medskip

\section{Quasi-equilibrium approximations to Einstein equations}
\label{s:EE}

Initial data for Cauchy evolutions of Einstein equations are constructed by solving the 
constraint equations. Under a conformal ansatz, constraints are cast as a scalar and a vector
elliptic equations.  In the XCTS approach \cite{CTS},
an additional elliptic equation for the (conformal) lapse follows from 
maximal slicing.
Part of the initial data must
be chosen freely. An approach to this
consists in solving the whole set of Einstein equations
under a quasiequilibrium approximation (e.g. 
\cite{ShiUryFri04,BonGouGra04,CooBau08}). We briefly review the approach in 
\cite{BonGouGra04}.
A fiducial time-independent flat metric  $f_{ab}$,  $\partial_t f_{ab}=0$, 
with connection ${\cal D}_a$
is introduced.
Then, a conformal decomposition of the data $(\gamma_{ab}, K^{ab})$ is
performed by choosing a conformal representative $\tilde{\gamma}_{ab}$ 
through the unimodular condition 
$ \det \tilde{\gamma}_{ab}=  \det f_{ab}$. We write
$  \gamma_{ab}  = 
  \Psi^4 \tilde{\gamma}_{ab},$ and 
$  K^{ab} = \Psi^{-4} \tilde{A}^{ab} + \frac{1}{3} K \gamma^{ab}$,
with $K=\gamma^{cd}K_{cd}$ 
and $\tilde{A}^{ab}$ 
\bea
\label{e:A_decomp}
  \tilde{A}^{ab} = \frac{1}{2 N}
  \left( \tilde{D}^a \beta^b + \tilde{D}^b \beta^a -
  \frac{2}{3} \tilde{D}_c \beta^c \tilde{\gamma}^{ab} +
  \partial_t \tilde{\gamma}^{ab} \right),
\eea
where $\tilde{D}_a$ is the Levi-Civita connection associated
with $\tilde{\gamma}_{ab}$. 
The  constrained evolution scheme in  \cite{BonGouGra04} gives rise to a mixed
elliptic-hyperbolic system whose elliptic subsystem is given by the following
equations.
The Hamiltonian constraint becomes an equation for
$\Psi$ whereas the momentum constraint translates into an equation for
$\beta^a$.
Maximal slicing and a Dirac-like 
gauge condition, namely  preservation in time of 
$K=0$ and ${\cal D}_c \tilde{\gamma}^{ca}=0$, are considered in 
\cite{BonGouGra04}. From $\dot{K}=0$ an elliptic equation for the lapse  $N$ follows.
The resulting XCTS-like elliptic subsystem on $\Psi$, $\beta^a$ 
and $N$ is formally expressed as
\bea
\label{e:elliptic}
L_\Psi \Psi = S_\Psi \ , \ L_\beta \beta^a = S^a_\beta \ , \ L_N N = S_N \ \ ,
\eea
with the  elliptic operators $L_\Psi$, $L_\beta$ and $L_N$ and the sources
$S_\Psi$,  $S^a_\beta$ and $S_N$ (cf. \cite{BonGouGra04}). 
The hyperbolic part consists of a wavelike equation on $\tilde{\gamma}^{ab}$ 
\bea
\label{e:wave_eq}
\frac{\partial^2 \tilde{\gamma}^{ab}}{\partial t^2} - \frac{N^2}{\Psi^4}
\tilde{\Delta}\tilde{\gamma}^{ab} 
- 2 {\cal L}_{\beta}\frac{\partial \tilde{\gamma}^{ab}}{\partial t} + {\cal
  L}_{\beta}{\cal L}_{\beta}\tilde{\gamma}^{ab} = S_{\tilde{\gamma}}^{ab} \ \ ,
\eea
with $S_{\tilde{\gamma}}^{ab}$ not depending on second derivatives
of $\tilde{\gamma}^{ab}$. The quasiequilibrium scheme
follows by setting in (\ref{e:wave_eq}) the values of 
$\partial_t \tilde{\gamma}^{ab}$ and 
$\frac{\partial^2 \tilde{\gamma}^{ab}}{\partial t^2}$ 
to appropriate {\em a priori} prescribed quantities. 
Then, Eqs. (\ref{e:wave_eq}) and (\ref{e:elliptic}) define an extended 
elliptic system.
In this brief paper we discuss neither outer boundary conditions nor bulk
quasiequilibrium prescriptions \cite{ShiUryFri04,BonGouGra04,CooBau08}, 
and focus on inner boundary conditions derived from IH structures when using
excision.

\medskip 

\subsection{NEH inner boundary conditions}
\label{s:NEHBC}

NEH conditions (\ref{e:NEH_BC}) 
and the gauge adaptation of the coordinate system to the
excision tube can be used to
fix four inner conditions in 
the elliptic subsystem (\ref{e:elliptic}). We
conformally rescale the relevant objects on 
${\cal S}$: $\tilde{q}_{ab} =  \Psi^{-4} q_{ab}$,
with connection $\tDS_a$, 
$\tilde{s}^a = \Psi^2 s^a$ and the 2+1 decomposition of the shift 
$\beta^a = \beta_\perp s^a + \beta_\parallel^a$, with  $\beta_\parallel^c s_c = 0$
and $\beta_\perp = \beta^c s_c$. Condition $\theta^{(\el)}=0$ in (\ref{e:NEH_BC}) translates
into
\bea
\label{e:BC_Psi}
\displaystyle 4\tilde{s}^c\tD_c\Psi + \tD_c\tilde{s}^c \Psi = \! &-& \! \Psi^{3}
\tilde{A}_{cd}\tilde{s}^c\tilde{s}^d + \frac{2}{3}\Psi^3K .
\eea
Coordinate adaptation to the horizon, namely $t^a = \ell^a 
+ \beta_\parallel^a$, and
the vanishing of $\sigma^{(\ell)}_{ab}$ 
in (\ref{e:NEH_BC}) [here we use
$\partial_t{\tilde{\gamma}}^{ab}=0$; cf. \cite{GouJar06} for general expressions] 
become conditions \cite{GouJar06}
\bea
\label{e:shearzero}
\displaystyle
\beta^\perp = N \ \ , \ \ \tDS^a \beta_\parallel^b + \tDS^b \beta_\parallel^a
    - (\tDS_c \beta_\parallel^c)\,  \tilde q^{ab} = 0 \ \ .
\eea 
In the XCTS context, a fifth boundary condition, generally interpreted as a condition 
on the lapse $N$,
is chosen arbitrarily to complete the elliptic system (\ref{e:elliptic}).

\medskip 

\subsection{IH inner boundary conditions.} 
\label{s:IHBC}

IH boundary conditions (\ref{e:IHconstraint}) represent three geometric 
conditions, to be set together with the NEH ones. As discussed above, 
under the $\ell^a$-normalization choice $f=N$,
IH equation (\ref{e:thetak}) on $\theta^{(k)}$ becomes a condition
on $N$. Therefore, all inner boundary conditions for the system (\ref{e:elliptic})
are determined. The only remaining freedom, necessary
to avoid degeneracies, is in the choice of the constant $\kappa_o$.

Regarding the elliptic system (\ref{e:wave_eq}) on $\tilde{\gamma}^{ab}$, we need 
to assess five boundary conditions 
(see \cite{CooBau08} for the first discussion
of this issue). 
The spatial gauge determines three of the
degrees of freedom associated with $\tilde{\gamma}^{ab}$ on ${\cal S}$
(cf. \cite{CooBau08,VasNov09} for two alternative perspectives on this, both based
on the use of Dirac-like gauges ${\cal D}_c\tilde{\gamma}^{ca}=0$).

IH condition (\ref{e:sigmak}) can then be used to set inner conditions for the 
remaining 2
degrees of freedom of $\tilde{\gamma}^{ab}$. Condition
(\ref{e:sigmak}) can be seen 
as a (nonlinear) Robin condition on two of the degrees of freedom
of $\tilde{\gamma}^{ab}$ (or $\tilde{q}^{ab}$). This follows from: 
i) the complete determination of  $H_{ab}$ from
joined NEH and IH conditions, ii) the writing 
$H_{ab} = \Psi^{2}(\tilde{H}_{ab} + 2 \tilde{s}^c \tilde{D}_c \ln \Psi\tilde{q}_{ab})$
[where $\tilde{H}_{ab}$
is the extrinsic curvature
of $({\cal S}, \tilde{q}_{ab})$ into $(\Sigma, \tilde{\gamma}_{ab})$], and iii)
the expression of $\tilde{H}_{ab}$ as the {\em radial} derivative 
$\tilde{H}_{ab} = \frac{1}{2}{q^c}_a {q^d}_b {\cal L}_{\tilde s} \tilde{q}_{cd}$.

The explicit form for the IH geometric boundary conditions is cumbersome, but it follows
straightforwardly from the expression of
(\ref{e:OmegakappaThetak}) in terms of conformal quantities
(we assume here $\partial_t\tilde{\gamma}^{ab}=0$, but cf. 
Eqs. (10.105) and (10.106) in \cite{GouJar06} for general expressions)
\bea
\label{e:conformal_Xi}
 && (\Theta^{(k)})^{ab}  =  \displaystyle 
    -\frac{1}{2N^2} 
    \left\{ (N + \beta^\perp) \Psi^{-2}
        \tilde{H}^{ab}
        + \frac{1}{2}\left[ \tD^a \beta^b_\parallel +\tD^b \beta^a_\parallel\right] \right.
\nonumber \\
 &&  \displaystyle   \left. + \left[ 2\Psi^{-2} \tilde{s}^c\tD_c\ln\Psi %[3mm]
      \displaystyle  
    + \frac{1}{3} \left( NK - \beta^\perp \Psi^{-2} \tilde H -
        \tilde{s}^c\tD_c (\beta^\perp \Psi^{-2})  \right. \right. \right.
\nonumber \\
&& \left. \left. \left.
- \tDS_c \beta^c_\parallel
        + \beta^c_\parallel \tilde{s}^d \tD_d \tilde{s}_c\right) \right] \, \tilde{q}^{ab}
        \right\} , \\
%\eea
%\bea
\label{e:conformal_Omega}     
&&    \Omega_a =  \displaystyle {\tDS}_a\ln N
    +\frac{\Psi^2}{2N} \left[ -(\beta^\perp \Psi^{-2})^2 \tilde{s}^c \tD_c \tilde{s}_a
    - \tDS_a (\beta^\perp \Psi^{-2}) \right. \nonumber\\
\displaystyle
&& \left.  + \tilde {H}_{ac}\beta^c_\parallel
    - \tilde{q}_{ac} \tilde{s}^d\tD_d \beta^c_\parallel \right] \ .
\eea

\medskip

\section{Discussion}
\label{s:discussion}

NEH conditions on an excised surface ${\cal S}$ determine three geometric conditions.
(Strongly) IH conditions determine three additional geometric conditions, up to 
a freely chosen constant (the surface gravity $\kappa_o$). In geometric terms, they
fully determine the {\em second fundamental form} ${\cal K}^c_{ab}$ of 
${\cal S}$ in a spacetime $({\cal M}, g_{ab})$. In a 3+1 description, 
this translates into the determination of both the extrinsic curvature $H_{ab}$ of ${\cal S}$ 
in $\Sigma$ and the projection onto ${\cal S}$ of the extrinsic curvature $K_{ab}$ 
of $\Sigma$ in ${\cal M}$. 

These conditions can be expressed in terms of initial data $(\gamma_{ab}, K^{ab})$
on a 3-slice $\Sigma$. This nontrivial feature permits their use 
as inner boundary conditions in the construction of initial data. 
We have illustrated this in the particular case of 
an elliptic system resulting from a 
conformal decomposition of Einstein equations under a quasiequilibrium ansatz.
Inner boundary conditions for a XCTS-like elliptic system (five equations) are determined 
from NEH conditions (three conditions), together with two additional ones: 
(i) the gauge adaptation of the coordinate
system to the excised tube, and (ii) the IH condition for $\theta^{(k)}$.
Inner boundary conditions for the elliptic system on the unimodular conformal metric $\tilde{\gamma}^{ab}$
follow from the spatial gauge (three conditions) and the traceless part of the 
IH conditions, i.e. the IH expression for $\sigma^{(k)}_{ab}$ (two conditions). 
Our main conclusion 
is the following: the full IH structure determines geometric conditions 
for a black hole in instantaneous equilibrium 
that fix (the physical) part of the inner boundary conditions
of the conformal metric $\tilde{\gamma}^{ab}$. In particular, they fully 
determine the shape of the excision surface ${\cal S}$ in $\Sigma$.

IH conditions may have some further physical interest.
Under maximal slicing and Dirac-like gauges, the only remaining freedom 
in the discussed system
is the choice of the constant $\kappa_o$.
This determines a one-parameter family of horizon foliations 
that fixes the inherent boost ambiguity in the IH description. 
On physical grounds, the fixing of the horizon-boost is expected to be related to a
(quasilocal) linear momentum of the black hole. The matching of the latter 
with the prescription following 
from a post-Newtonian expansion could eventually be used to fix a preferred
value of the parameter $\kappa_o$.

There is a number of important caveats in the present approach. 
First, we have only addressed issues of geometric nature, 
with no reference whatsoever to the analytic well-posedness of the considered 
elliptic boundary problem. Such an analysis 
is crucial to assess the actual validity of the IH conditions here discussed. 
The study in Ref. \cite{VasNov09} is illustrative in this respect, since it shows
a particular example (namely the single black hole case)
where no inner boundary conditions for $\tilde{\gamma}^{ab}$ need to be specified.
Though that example is fully compatible with the present IH proposal (the constructed
Kerr data fulfill the inner IH conditions), it also points out the need
of a general analytic study, something beyond the scope of the present geometric discussion. 
Second, the physical convenience of using (strongly) IH in the 
initial data construction can be called into question.
They may represent too stringent conditions in certain realistic astrophysical 
situations
where the use of NEH boundary conditions could prove to be enough.
Third, the implementation of (strongly) IH conditions
can be challenging from a numerical point of view.
In the context of the last two caveats, the free ({\em effective}) inner boundary
conditions for $\tilde{\gamma}^{ab}$ proposed in \cite{CooBau08}
represent an interesting alternative,
at least in {\em generic} cases. 
In contrast with IH conditions, prescribing the shape of ${\cal S}$,
conditions in \cite{CooBau08} fix the (conformal) intrinsic geometry of ${\cal S}$.
Technically,
they are considerably simpler than IH conditions. 
All these issues must be assessed numerically.

The author acknowledges E. Gourgoulhon, P. Grandcl\'ement, 
B. Krishnan, J. Novak, M. S\'anchez, 
J. A. Valiente-Kroon and N. Vasset for enlightening discussions. 
The author has been supported by the Spanish MICINN 
under project FIS2008-06078-C03-01/FIS and 
Junta de Andaluc\'{\i}a under projects FQM2288, FQM219.

\end{document}